\documentclass[12pt]{article}
\input{epsf}

\font\bm=cmmib10 at 10pt
\font\bms=cmmib10 at 7pt \textfont9=\bm \scriptfont9=\bms
\usepackage{amsfonts}
\mathchardef\balpha= "790B
\mathchardef\bbeta= "790C
\mathchardef\bTheta= "7902
\mathchardef\bzeta= "7910
\mathchardef\bOmega= "790A
\mathchardef\bGamma= "7900
\mathchardef\bDelta= "7901
\mathchardef\bPhi= "7908
\mathchardef\bphi= "791E
\mathchardef\bomega= "7921
\mathchardef\bxi= "7918
\mathchardef\bet= "7911
\mathchardef\brho= "791A
\mathchardef\btau= "791C
\mathchardef\bmu= "7916
\mathchardef\bvarpi= "7924

\def \lvec{(\kern-.26em(}
\def\pmb#1{\setbox0=\hbox{#1}%
\def \lvec{(\kern-.26em(}
\kern-.025em\copy0\kern-\wd0
\kern.05em\copy0\kern-\wd0
\kern-.025em\raise.0433em\box0 }
\mathchardef\btheta= "7912

\def\XXint#1#2#3{{\setbox0=\hbox{$#1{#2#3}{\int}$}
     \vcenter{\hbox{$#2#3$}}\kern-.5\wd0}}

\usepackage{amsmath}
\usepackage{authblk}
\usepackage{url}

\usepackage{graphicx}
\usepackage{dcolumn}

\begin{document}

\title{Sunlight, the Bond Albedo, CO$_2$, and Earth's Temperature}
\author[1] {R. Louw}
\author[2]{ W. A. van Wijngaarden}
\author[3] {W. Happer}

\affil[1]{Chemical Engineer, Stokesley, England}
\affil[2]{Department of Physics and Astronomy, York University, Canada}
\affil[3]{Department of Physics, Princeton University, USA}
\renewcommand\Affilfont{\itshape\small}
\date{\today}
\maketitle

\begin{abstract}
The main determinants of Earth's absolute surface temperature, $T$, are the {\it solar constant}, $S$, the {\it Bond albedo}, A,   and the {\it effective emissivity} for thermal radiation, $\epsilon$. In this note we assume that the value of the effective emissivity, $\epsilon =\epsilon(C)$, is determined by the atmospheric concentration $C$ of CO$_2$. We show that the solar constant  is most important, the albedo is second, and the CO$_2$ concentration is a distant third.
\end{abstract}

The surface-averaged  heating flux, $Z_{\rm s}$, of the Earth by sunlight can be written as
\begin{equation}	
Z_{\rm s}=\frac{1}{4}S(1-A).
\label{c6}
\end{equation}
Here 
\begin{equation}	
S= 1361\hbox{ W m$^{-2}$},
\label{c8}
\end{equation}
 is the solar constant\,\cite{S}, the yearly averaged power passing through a square meter area that is normal to the direction of a ray from the center of the Sun to an observation point on Earth's orbit. The solar constant is not really constant in time, and it varies by at least a few parts per thousand over a  11-year solar cycle. It is possible that there are larger, longer-period variations of $S$ that have not yet been characterized, but which have a signficant influence on climate.

Taking the radius of the Earth to be $r$, the factor of $1/4$  of (\ref{c6}) comes from the ratio, $1/4= \pi r^2/4\pi r^2$ of the area $\pi r^2$ of the circular cross section of the Earth that is illuminated by sunlight to the total surface area, $4\pi r^2$. The Bond albedo $A$ of (\ref{c6}) is the fraction of solar radiation incident on the Earth that is reflected to space and not converted to heat by absorption in the atmosphere or surface\,\cite{Bond}. The observed value of the Bond albedo is approximately 
\begin{equation}	
A=0.30.
\label{c10}
\end{equation}
Contemporary small variations in the Bond albedo are mainly due to variations in cloud cover, and to a lesser extent, snow and ice cover.

The upward flux  $Z_{\rm t}$ of thermal radiation to space can be written as
\begin{equation}	
Z_{\rm t}=\epsilon\sigma T^4.
\label{c2}
\end{equation}
Here 
\begin{equation}	
\sigma =5.67\times 10^{-8} \hbox{ W m}^{-2}\hbox{  K}^{-4},
\label{c2a}
\end{equation}
is the Stefan-Boltzmann constant. 
We can think of (\ref{c2}) as the definition of the effective emissivity, $\epsilon$. For a hypothetical, isothermal, black Earth, with no greenhouse gases, the effective emissivity would be $\epsilon = 1$. For the real Earth the ``local" effective emissivity for the local surface temperature  is $\epsilon<1$ for most of Earth's surface,  mainly because the upward flux from the warm surface is replaced by weaker upward flux from greenhouse gases and clouds in the cold upper atmosphere. An exception is wintertime, cloud-free Antarctica, where there is normally a large temperature inversion.  There, more radiation reaches outer space from  CO$_2$ and H$_2$O emissions in the relatively warm upper atmosphere than is emitted at the same frequencies by the icy surface. So we often find $\epsilon>1$ for winter Antarctica. 

In (\ref{c2}) we will take the average surface temperature of the Earth to be
\begin{equation}	
T= 288.7\hbox{ K}.
\label{c3}
\end{equation}
In thermal equlibrium, the radiative cooling (\ref{c2}) of Earth by long-wave infrared radiation, will be equal to the solar heating (\ref{c6}) and we can write
\begin{equation}	
\frac{1}{4}S(1-A)=\epsilon\sigma T^4.
\label{c14}
\end{equation}
An effective emissivity, averaged over the entire surface of the Earth, can be found by solving  (\ref{c14}) with  the values of $S$, $A$, $\sigma$ and $T$ from (\ref{c8}), (\ref{c10}), (\ref{c2a}) and (\ref{c3}) to find
\begin{equation}	
\epsilon= 0.60.
\label{c4}
\end{equation}

Taking natural logarithms of both sides of (\ref{c14}) we find
\begin{equation}	
-\ln 4 +\ln S+\ln(1-A)=\ln \epsilon+\ln \sigma +4\ln T.
\label{c16}
\end{equation}
Suppose some change in the concentrations of greenhouse gases or cloud cover, or the solar constant  causes the variables $S$, $A$, $\epsilon$ and $T$ to change by small increments  $dS$, $dA$, $d\epsilon$ and $dT$. We assume that there is enough time after the changes to reestablish thermal equilibrium so (\ref{c14}) and (\ref{c16}) remain valid. Then we can take differentials of both sides of (\ref{c16}) to find
\begin{equation}	
\frac{dS}{S}-\frac{dA}{1-A}=\frac{d \epsilon}{\epsilon}+4\frac{dT}{T}.
\label{c18}
\end{equation}
Solving (\ref{c18}) for the relative change in temperature we find
\begin{eqnarray}	
\frac{dT}{T}&=&\frac{1}{4}\left(\frac{dS}{S}-\frac{dA}{1-A}-\frac{d \epsilon}{\epsilon}\right)\nonumber\\
&=&\frac{1}{4}\left(d\ln S-\frac{A}{1-A}d\ln A-d \ln \epsilon\right),
\label{c20}
\end{eqnarray}
or
\begin{eqnarray}	
d\ln T&=&\left(\frac{\partial \ln T}{\partial \ln S}\right) d \ln S+\left(\frac{\partial \ln T}{\partial \ln A} \right) d \ln A 
+\left(\frac{\partial \ln T}{\partial \ln \epsilon}\right)  d \ln\epsilon \nonumber\\
&=&(0.2500) d\ln S-(0.1071) d\ln A-(0.2500)d \ln \epsilon.
\label{c21}
\end{eqnarray}
The numerical coefficient of $d\ln A$ in the bottom line of (\ref{c21}) comes  from using the value $A=0.30$ from (\ref{c10})  in (\ref{c20}).  If we assume that the effective emissivity $\epsilon=\epsilon(C)$  depends only on the CO$_2$ concentration $C$, we can write (\ref{c21}) as
\begin{eqnarray}	
d\ln T
&=&(0.2500) d\ln S-(0.1071) d\ln A-(0.2500)\left(\frac{d\ln\epsilon}{d\ln C}\right)d \ln C.
\label{c22}
\end{eqnarray}

To find the factor $d\ln \epsilon/d\ln C$ of (\ref{c22}) we recall\,\cite{sg} that for clear skies in temperate latitudes 
 the thermal flux (\ref{c2}) can be written in terms of a CO$_2$ induced forcing $F(C)$
\begin{eqnarray}	
Z_{\rm t}(C)&=&Z_{\rm t}(0)-F(C)\nonumber\\
&=&277  \hbox { W m}^{-2} \quad \hbox{for} \quad C=400 \hbox{ ppm}.
\label{c30}
\end{eqnarray}
A simple empirical formula\,\cite{sg} that gives a good fit to the computer-calculated forcing of reference\,\cite{WH} is
\begin{eqnarray}
F(C)&=&\Delta F\log_2(1+C/C_0)\nonumber\\
&\approx &\Delta F\log_2(C/C_0) \quad\hbox{if}\quad C\gg C_0,
\label{c32}
\end{eqnarray}
where
\begin{equation}
\Delta F= 3\, \hbox { W m}^{-2}.\label{c34}
\end{equation}
and 
\begin{equation}
C_0 = 0.391 \hbox{ ppm}.\label{c46}
\end{equation}
From (\ref{c2}) and (\ref{c30}), and for $C\gg C_0$,  we can write
\begin{equation}	
\epsilon\sigma T^4 = Z_{\rm t}(0)-\Delta F\log_2(C/C_0). 
\label{c36}
\end{equation}
We differentiate the left side of (\ref{c36}) with respect to $C$, for constant $T$, to find
\begin{eqnarray}	
d\left(\epsilon\sigma T^4\right) &=& \sigma T^4 d\epsilon\nonumber\\
&=&\epsilon\sigma T^4 d\ln \epsilon\nonumber\\
&=&Z_{\rm t}(C) d \ln \epsilon.
\label{c38}
\end{eqnarray}
Differentiating the right side (\ref{c36}) in like manner, we find
\begin{eqnarray}	
d \left[ Z_{\rm t}(0)-\Delta F\log_2(C/C_0)\right]&=&-\Delta F d\log_2 C\nonumber\\
&=&-\log_2 \!e\, \Delta F d\ln C.
\label{c40}
\end{eqnarray}
Equating (\ref{c38}) and (\ref{c40}) we find

\begin{eqnarray}	
\left(\frac{d\ln \epsilon}{d\ln C}\right) &=&-\frac{\log_2 e\, \Delta F}{Z_{\rm t}(C)}\nonumber\\
&=&-\frac{1.4427 \times 3}{277}\nonumber\\
&=&-0.01562
\label{c42}
\end{eqnarray}
Using (\ref{c42}) in (\ref{c22}) we find
\begin{eqnarray}	
d\ln T&=&\left(\frac{\partial \ln T}{\partial \ln S}\right) d \ln S+\left(\frac{\partial \ln T}{\partial \ln A}\right) d \ln A 
+\left(\frac{\partial \ln T}{\partial \ln C}\right) d \ln C \nonumber\\
&=&(0.2500) d\ln S-(0.1071) d\ln A+(0.003906) d \ln C.
\label{c44}
\end{eqnarray}
\begin{figure}[t]
\includegraphics[height=80 mm, width=1
\columnwidth]{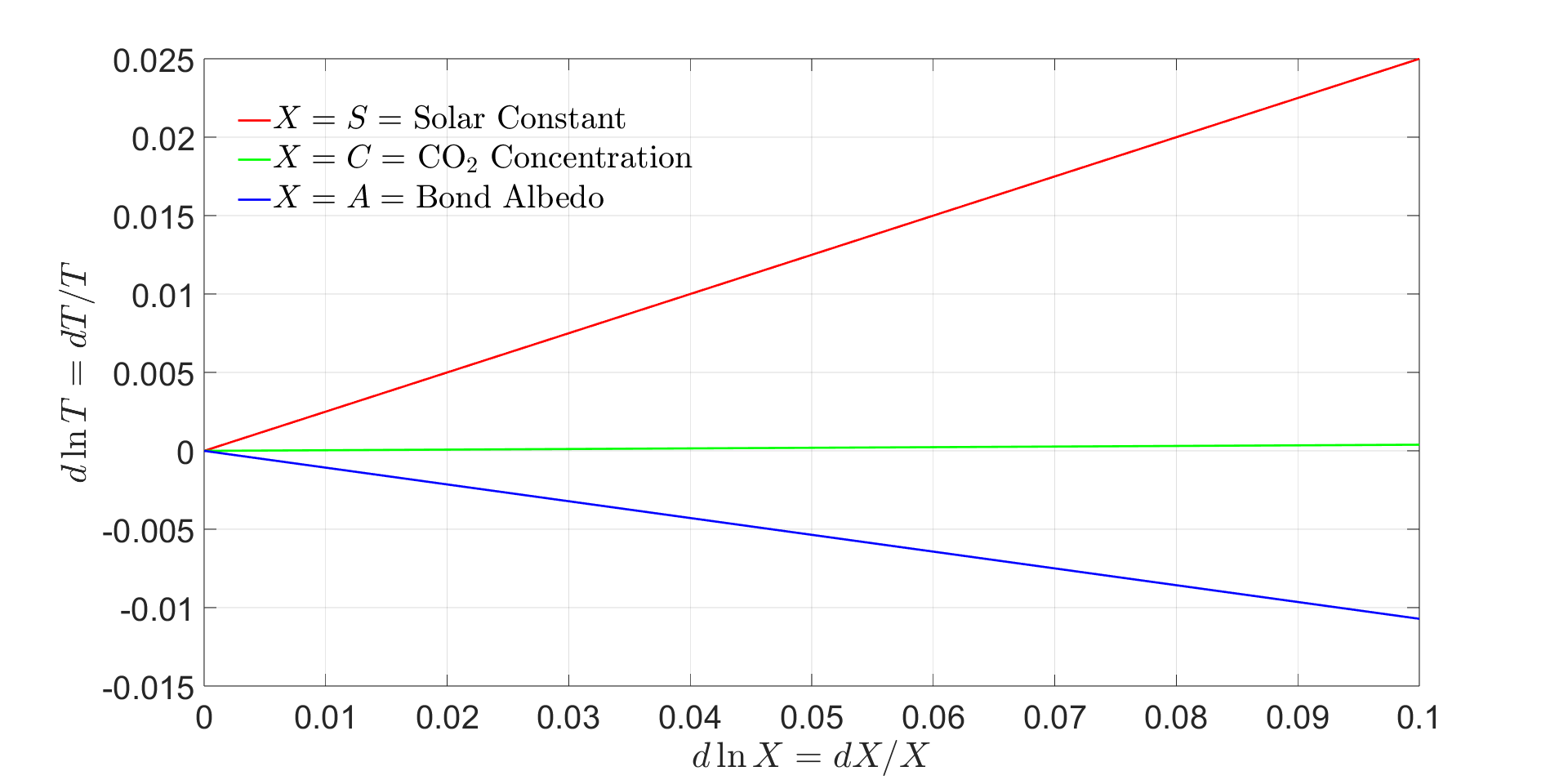}
\caption {The relative changes $d \ln T=dT/T$ of Earth's average, absolute surface temperature $T$ caused by relative changes   $d \ln S=d S/S$, $d \ln C=d C/C$ and  $d \ln A=d A/A$ of the solar constant $S$, the CO$_2$ concentration $C$, and the Bond albedo $A$. Relative changes of CO$_2$ have a much smaller effect on temperature than changes by the same relative amount of the solar constant $S$ or the Bond albedo $A$.
\label{TSCA}}
\end{figure}

From (\ref{c44}) we see that 
\begin{equation}	
\left(\frac{\partial \ln T}{\partial \ln C}\right)=0.003906,
\label{c46}
\end{equation}
and
\begin{eqnarray}	
\left(\frac{\partial \ln T}{\partial \ln A}\right)&=&-0.1071\nonumber\\
&=&-27.4\left(\frac{\partial \ln T}{\partial \ln C}\right),
\label{c48}
\end{eqnarray}
where the value of $\partial \ln T/\partial \ln C$ in the second line is given by (\ref{c46}).
According to (\ref{c48}), a relative change $d \ln A = d A/A$ of the Bond albedo
 $A$  causes a 27.4 times larger relative change (of opposite sign) in temperature $d\ln T=d T/T$, than a relative change  of the same magnitude, $d \ln C = d C/C $, of the CO$_2$ concentration $C$.

Eq. (\ref{c44}) also implies that
\begin{eqnarray}	
\left(\frac{\partial \ln T}{\partial \ln S}\right)&=&0.2500\nonumber\\
&=&64\left(\frac{\partial \ln T}{\partial \ln C}\right).
\label{c50}
\end{eqnarray}
According to (\ref{c50}), a relative change $d \ln S = d S/S$ of the solar constant
$S$ causes a 64 times larger relative change of temperature $d\ln T=d T/T$, than a relative change  of the same magnitude, $d \ln C = d C/C$, of the CO$_2$ concentration $C$.

In summary,  Fig. \ref{TSCA}, (\ref{c48}) and (\ref{c50}) show that the most important influence on Earth's surface temperature is the solar constant $S$, followed closely by the Bond albedo $A$. The CO$_2$ concentration $C$ has much less influence. For the same relative changes $dC/C$ of the atmospheric CO$_2$ concentrations, relative changes $dS/S$ or $dA/A$ cause relative warmings $dT/T$ that are 64 and 27 times larger, respectively. As discussed in \cite{sg} and \cite{WH}, this is because of the heavy saturation of the radiative forcing of CO$_2$. 

We have made many simplifying assumptions to keep this note as brief as possible. For example, we have characterized the Earth with a single characteristic temperature $T$ or a single temperature change $dT$, as is commonly done for discusions of climate.  Clouds probably diminish the clear-sky forcing changes due to changes in  CO$_2$ concentration, by about 30\%. So the importances of the solar constant $S$ and the albedo $A$ relative to the CO$_2$ concentration $C$ are probably greater than indicated by (\ref{c48}) and (\ref{c50}), which were estimated for clear-sky conditions.


\begin{thebibliography}{99} 
\bibitem{S} The Solar Constant\\
\url{https://en.wikipedia.org/wiki/Solar_constant}
\bibitem{Bond} The Bond Albedo\\
\url{https://en.wikipedia.org/wiki/Bond_albedo}
\bibitem{sg}Saturation Graphics,\\ \url{https://co2coalition.org/publications/saturation-graphics/}
\bibitem{WH} W. A. van Wijngaarden and W. Happer, {\it Dependence of Earth's Thermal Radiation on Five Most Abundant Greenhouse Gases}, Atmos. \& Oceanic Phys., arXiv: 2006.03098 (2020).
\end{thebibliography}
\end{document}